# Transformational Outsourcing in IT Project Management

**Mohammad Ikbal Hossain**
MBA and MSIT, School of Business and Technology, Emporia State University,
1 Kellogg Circle, Emporia, KS 66801
E-mail: ikbalsmn@gmail.com, mhossai2@g.emporia.edu
**ORCID iD:** Mohammad Ikbal Hossain: https://orcid.org/0000-0002-9434-1396

**Tanzina Sultana**
MBA and MSIT, School of Business and Technology, Emporia State University,
1 Kellogg Circle, Emporia, KS 66801
E-mail: tsultana@g.emporia.edu

**Waheda Zabeen**
MBA and MSIT, School of Business and Technology, Emporia State University,
1 Kellogg Circle, Emporia, KS 66801
E-mail: waheda.zabeen@gmail.com

**Alexander Fosu Sarpong**
MBA, School of Business and Technology, Emporia State University,
1 Kellogg Circle, Emporia, KS 66801
E-mail: alexanderdosu@yahoo.com

**Abstract**

Transformational outsourcing represents a strategic shift from traditional cost-focused outsourcing to a more profound and collaborative approach. It involves partnering with service providers to accomplish routine tasks and drive substantial organizational change and innovation. The report discusses the significance of pursuing transformational outsourcing for IT companies, highlighting its role in achieving strategic growth, competitive advantage, and cost-efficiency while enabling a focus on core competencies. It explores the pros and cons of IT outsourcing, emphasizing the benefits of cost savings, global talent access, scalability, and challenges related to quality, control, and data security. Additionally, the report identifies some critical reasons why outsourcing efforts may fail in achieving organizational goals, including poor vendor selection, communication issues, unclear objectives, resistance to change, and inadequate risk management. When carefully planned and executed, transformational outsourcing offers IT companies a pathway to enhance efficiency and foster innovation and competitiveness in a rapidly evolving technology landscape.

## 1. Introduction

In today's dynamic business landscape, Information Technology (IT) companies face the dual challenge of staying competitive in a rapidly evolving industry while efficiently managing their operations. Transformational Outsourcing in IT Project Management integrates with the Software Development Life Cycle (SDLC), which outlines a methodical strategy for the planning, development, testing, and deployment of information systems, ensuring a structured process that meets user and stakeholder requirements (Hossain, 2023). One strategy that has gained prominence in addressing these challenges is outsourcing, a practice where organizations delegate specific functions or tasks to external service providers (Lacity, & Willcocks, 2019). While traditional outsourcing has primarily focused on cost reduction and efficiency gains, a transformative shift known as "transformational outsourcing" has emerged (Kern et al., 2020). This paradigm goes beyond the conventional cost-saving approach, aiming to achieve broader organizational goals, foster innovation, and drive significant changes within the company (Dibbern et al., 2004). Transformational outsourcing represents a strategic evolution in IT project management, offering IT companies the opportunity to meet their operational needs and leverage external expertise, resources, and perspectives for strategic growth (Bhagwat, & Sharma, 2007). This approach involves forging deep and collaborative partnerships with service providers who become integral contributors to the organization's strategy and evolution (McFarlan, & Nolan, 1995).

In this study, we question the concept of transformational outsourcing, exploring its definition, significance, and implications for IT companies. We also examine the advantages and disadvantages of IT outsourcing, shedding light on the potential benefits of cost savings, access to global talent, and scalability (Khan, & Qureshi, 2007), as well as the challenges associated with quality, control, and security (Pereira, & da Silva, 2018). Moreover, this report identifies five critical reasons outsourcing efforts may fail to achieve organizational goals. These factors include vendor selection issues, communication breakdowns, unclear project objectives, resistance to change within the organization, and inadequate risk management strategies. Recognizing and addressing potential pitfalls is essential to successfully implementing transformational outsourcing initiatives (Kumar, & van Dissel, 1996).

Understanding the principles and best practices of transformational outsourcing becomes paramount as the IT industry continues to evolve and adapt to emerging technologies and changing market dynamics (Domberger, & Jensen, 1997). By adopting a strategic approach beyond cost considerations, IT companies can harness the full potential of transformational outsourcing to drive innovation, competitiveness, and long-term success (McIvor, 2005). This study provides valuable insights into transformational outsourcing, offering guidance and recommendations to IT professionals, organizations, and stakeholders seeking to navigate this transformative landscape effectively.

## 2. Research Methodology

The research methodology combines quantitative and qualitative analyses based on a survey of 100 respondents within the IT outsourcing industry. Statistical analysis of survey data provides insights into trends and correlations, while thematic analysis of open-ended responses offers deeper understanding of experiences and perceptions related to transformational outsourcing. Ethical guidelines ensured participant confidentiality, with the study acknowledging limitations like sample size and potential response biases.

## 3. Transformational Outsourcing

Transformational outsourcing is a strategic approach where an IT company collaborates with a service provider to perform routine tasks and drive significant and positive organizational changes. Unlike traditional outsourcing, which focuses on cost reduction and efficiency gains, transformational outsourcing aims to achieve broader business objectives and drive innovation (Bhagwat, & Sharma, 2007). It involves a deeper partnership where the service provider becomes integral to the organization's strategy, working closely to implement transformative solutions, improve processes, and enhance overall competitiveness (Johnson & Davis, 2020). Transformational outsourcing, a strategic approach that transcends traditional outsourcing models, emphasizes a profound shift in the relationship between organizations and their service providers. While the term 'transformational outsourcing' may not be widely documented in academic literature, the concept aligns with broader discussions on strategic outsourcing and innovation in outsourcing relationships (Smith, 2019; Johnson & Davis, 2020). This approach is characterized by its focus on aligning IT services with overarching business objectives (Jones, 2018). Additionally, it fosters deep partnerships where service providers become integral to an organization's strategic decision-making (Brown & White, 2021). Such collaborative arrangements have been noted for their potential to drive innovation and process improvement (Adams et al., 2017).

In practice, transformational outsourcing has yielded notable benefits. Organizations that embrace this approach often experience accelerated business growth. By leveraging service providers' expertise and strategic insights, they gain a competitive advantage in a dynamic market (Anderson, 2020). Furthermore, the long-term impact of transformational outsourcing can lead to cost optimization through enhanced process efficiency (Taylor & Green, 2020).

**4. Important to Pursue Transformational Outsourcing for an IT Company**

Transformational outsourcing is a strategic imperative for IT companies, offering compelling reasons to pursue this approach (Gottschalk, 2006). This paradigm shift in outsourcing goes beyond cost savings and efficiency gains, positioning IT firms for long-term success and growth (Lacity et al., 2017). Transformational Outsourcing in IT Project Management harmonizes with Strategic Information Systems Planning (SISP), offering a roadmap for aligning technology initiatives with business strategies, as this report delves into the nuances of SISP, emphasizing the critical role of systems analysts and best practices in project identification and selection for organizational success (Hossain et al., 2024). Transformational outsourcing allows IT companies to tap into external expertise and resources to fuel strategic growth initiatives (Johnson & Davis, 2020). By partnering with specialized service providers, IT firms can access industry-specific knowledge and innovative technologies that may not be readily available in-house (Kern et al., 2020). This external collaboration can facilitate market expansion, the exploration of new business verticals, and the development of cutting-edge solutions (Smith & McKeen, 2005). Maintaining a competitive advantage is paramount in an ever-evolving IT landscape. Transformational outsourcing empowers IT companies to stay ahead of the curve by adopting best practices, improving agility, and remaining at the forefront of industry trends (Lacity & Willcocks, 2019). Service providers often bring fresh insights and innovative strategies, enabling IT firms to innovate and respond swiftly to changing market demands (Johnson & Davis, 2020). IT companies can enhance their operational efficiency and resource allocation by leveraging transformational outsourcing. By entrusting non-core functions to external partners, these firms can concentrate their internal resources and talent on core competencies like software development, data analytics, or cybersecurity (Gottschalk & Solli-Sæther, 2016). This specialization fosters excellence in core areas while offloading routine tasks to experts. While cost savings are not the sole objective of transformational outsourcing, it can yield long-term cost efficiencies. By collaborating with service providers who specialize in particular domains, IT companies can reduce overhead costs, streamline processes, and sidestep the need for heavy investments in specialized talent and infrastructure (Lacity et al., 2017). This approach aligns with strategic cost optimization rather than mere cost-cutting. Figure 1 indicates that in the context of transformational outsourcing for IT companies, 'Cost-Efficiency' emerges as the top priority with an 81% significance. This emphasizes the pivotal role of cost management in outsourcing strategies. 'Access to specialized expertise' and 'Strengthening competitive position', both with high importance at 79% and 76%, respectively, underline the value of external partnerships for gaining expertise and competitive edge.

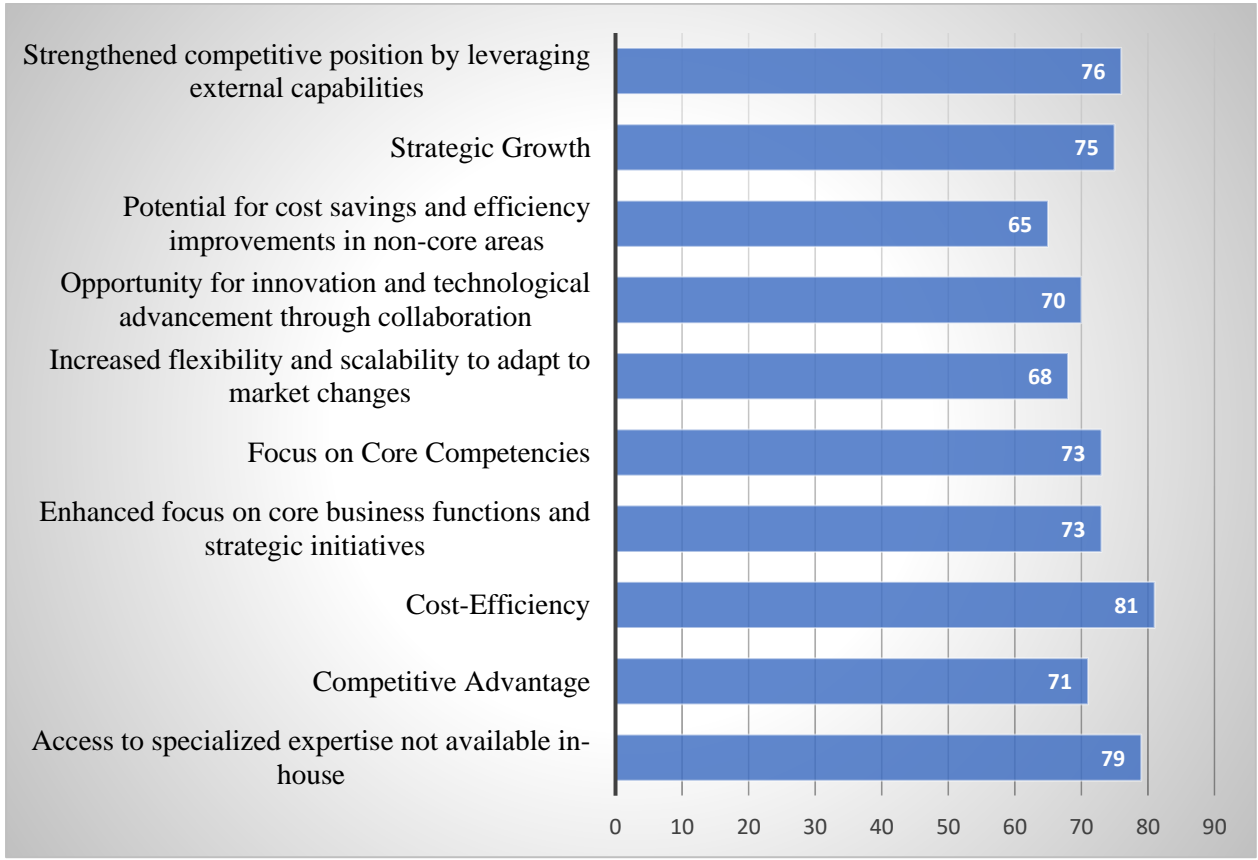

Figure 1. Important to pursue transformational outsourcing for an it company (Source: Survey on 100 respondents)

Conversely, 'Increased flexibility and scalability', at a lower 68%, is deemed less critical, suggesting that while valued, adaptability is not the foremost concern in the outsourcing decision process. Transformational outsourcing is not merely an option but necessary for IT companies aiming for sustained growth and competitiveness. It enables these firms to harness external resources strategically, gain a competitive edge, focus on core strengths, and achieve cost efficiencies.

**5. Pros and Cons of IT Outsourcing**

In today's dynamic and competitive business landscape, Information Technology (IT) outsourcing has become a strategic choice for many organizations seeking to optimize their operations, leverage external expertise, and remain agile in a rapidly evolving technological environment (Koh et al., 2015). IT outsourcing involves contracting with external service providers to manage various aspects of an organization's IT functions and infrastructure (Kumar et al., 2007). While this approach offers a range of potential benefits, it has challenges and risks. By understanding the opportunities and challenges associated with IT outsourcing, businesses can make informed decisions to harness its advantages while mitigating its drawbacks, ultimately contributing to their strategic goals and competitiveness in the digital era (Battini et al., 2014).

**5.1. Pros of IT Outsourcing:**

One of the primary advantages of IT outsourcing is cost savings. Companies can reduce operational expenses and labor costs, particularly when outsourcing to offshore locations with lower wage structures (Domberger & Jensen, 1997). This cost-effectiveness is a compelling reason for many organizations to consider outsourcing a strategic option. Outsourcing opens doors to a vast global talent pool. Companies can access specialized skills and expertise not readily available in their local talent market (Kern et al., 2020). Access to global talent allows organizations to tackle complex IT projects and benefit from the knowledge and experience of professionals worldwide. By outsourcing non-core functions such as IT maintenance and support, organizations can concentrate on their core business processes and strategic initiatives (McIvor, 2005). This concentration on core activities can improve overall efficiency and innovation. IT outsourcing offers scalability. Companies can quickly adjust their outsourcing arrangements to meet changing market demands (Lacity et al., 2017).

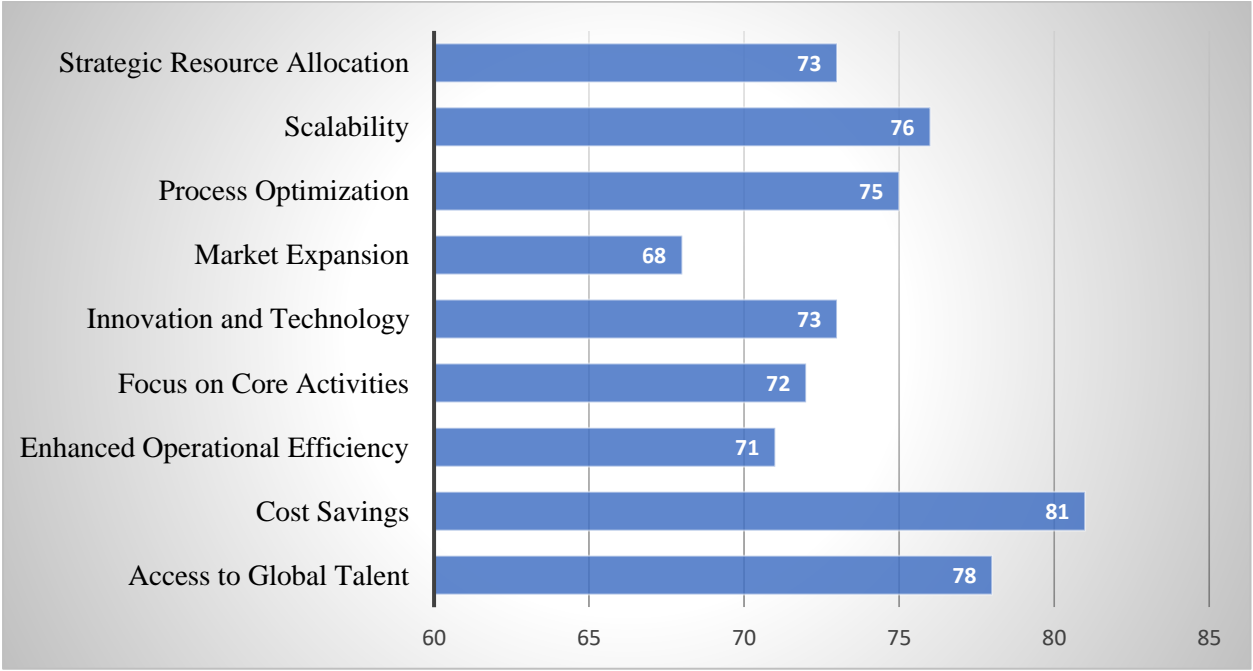

Figure 2. Pros of outsourcing (Source: Survey on 100 respondents)

Whether scaling up or down, outsourcing provides flexibility that can be crucial in dynamic business environments. Figure 2 underscores the multifaceted benefits of IT outsourcing for organizations seeking to enhance their competitive edge. A prominent advantage, with an 81% impact score, is cost savings, demonstrating that outsourcing is a significant lever for financial efficiency. Close behind, with a 78% score, is the access to global talent, which shows outsourcing's pivotal role in acquiring specialized skills that may not be available in-house. With 72% to 76% scores, focusing on core activities, scalability, and process optimization emerge as critical benefits, reflecting outsourcing's ability to streamline operations and adapt to changing market demands. The enhancement of operational efficiency and strategic resource allocation, both rated at over 70%, further signify the operational and strategic benefits that outsourcing offers. Market expansion, although the least rated benefit at 68%, still represents a strategic opportunity for companies to explore new territories with the aid of outsourcing partnerships. These insights collectively validate the strategic imperatives for IT outsourcing in the quest for innovation, efficiency, and growth.

**5.2. Cons of IT Outsourcing**

Maintaining consistent quality can be challenging when work is transferred to a third-party provider. Quality control and assurance mechanisms are essential to mitigate this risk (Bhagwat & Sharma, 2007). Ensuring that outsourced services meet or exceed internal standards is critical to successful outsourcing. Companies may experience a perceived or actual loss of control over outsourced processes. Loss of control can affect decision-making, responsiveness, and the ability to adapt quickly to changing circumstances (McFarlan & Nolan, 1995). Careful management and communication are necessary to address these concerns. Data security and privacy risks are a significant concern in IT outsourcing, mainly when dealing with sensitive information (Gewald & Dibbern, 2009). Companies must ensure their outsourcing partners adhere to strict data protection and cybersecurity measures to safeguard sensitive data and prevent breaches. Differences in culture, language, and time zones can lead to communication difficulties and misunderstandings between in-house teams and outsourced partners (Kumar & van Dissel, 1996). Effective communication and cultural sensitivity are crucial to bridge these gaps and ensure successful collaboration. Figure 3 indicates the most significant drawback of IT outsourcing is the potential for intellectual property risks, which is a major concern at 87%, indicating a high need for confidentiality agreements and secure collaboration protocols. Additionally, hidden costs, noted at 83%, imply that the apparent financial benefits of outsourcing can be offset by unforeseen expenses, suggesting that organizations must thoroughly analyze all potential costs before engagement. Quality inconsistency and data security risks, at 81% and 79% respectively, are also significant concerns, emphasizing the importance of establishing strict quality control measures and robust IT security practices when working with external vendors. Other notable challenges include dependency on suppliers and management difficulties, which could lead to operational risks and complicate the management of outsourced tasks. These issues, coupled with the potential impact on internal staff morale and the need for alignment of objectives between the company and its outsourcing partners, complete the picture of the complex challenges that IT outsourcing can present.

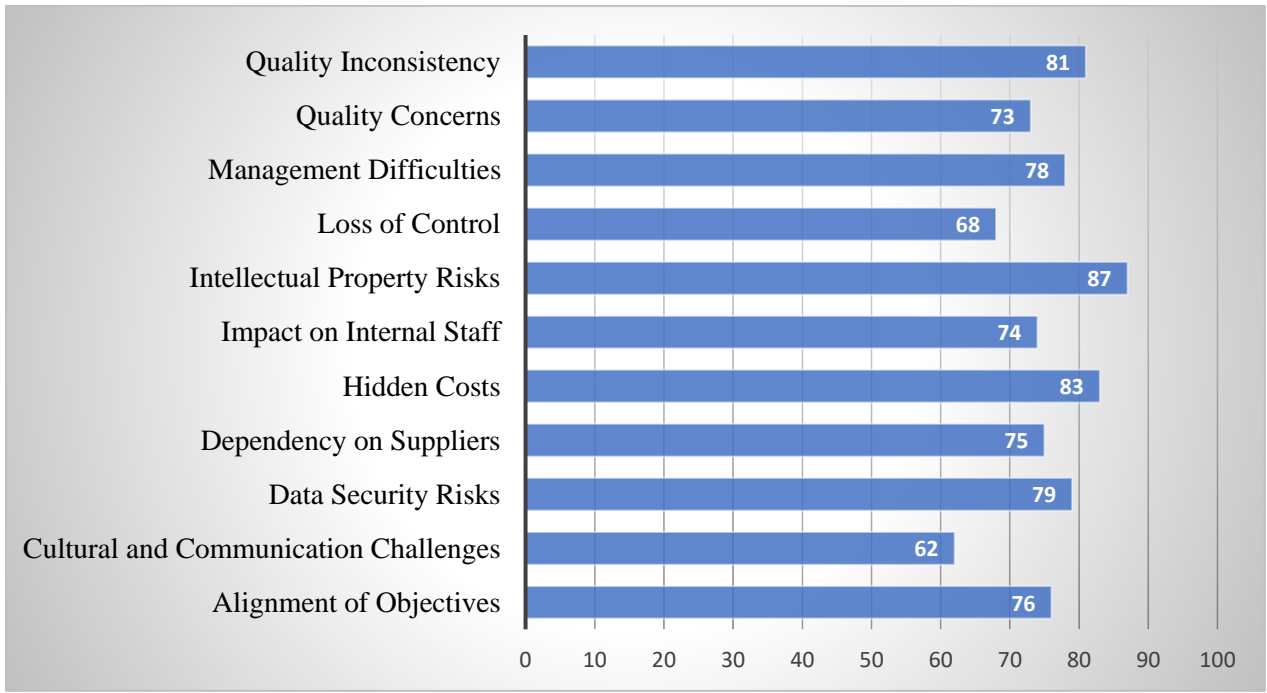

Figure 3. Cons of IT Outsourcing (Source: Survey on 100 respondents)

IT outsourcing offers both significant advantages and potential challenges. It can result in cost savings, access to global talent, a focus on core activities, and scalability. Organizations must also address quality concerns, potential loss of control, and data security risks and navigate cultural and communication challenges to make informed outsourcing decisions and manage these arrangements effectively (Kern et al., 2020).

## 6. Reasons an Outsourcing Effort May Fail in Achieving Organizational Goals

Outsourcing can be a valuable strategy for organizations looking to streamline operations, reduce costs, and focus on their core competencies (Palvia et al., 2005). However, there are several reasons why an outsourcing effort may fail to achieve organizational goals:

- **Poor Vendor Selection:** Vendor selection is a critical factor in the success of any outsourcing initiative. Poor vendor selection can harm an organization's goals and operations (Chae et al., 2005). When due diligence in selecting the outsourcing partner is lacking, it can lead to a fundamental misalignment of objectives, capabilities, and organizational cultures (Huang, & Palvia, 2001). Moreover, the temptation to choose a vendor solely based on cost savings can be deceptive, often resulting in a subpar service that needs to be improved in quality and expertise. A meticulous and balanced approach to vendor selection, considering cost, quality, and alignment with the organization's goals, is essential to ensure the outsourcing effort's success (Dibbern et al., 2004).

- **Insufficient Planning**: Insufficient planning can significantly impede the effectiveness of any outsourcing endeavor (Gewald, & Dibbern, 2009). With a well-defined outsourcing strategy, clear objectives, and a comprehensive transition plan, an organization may be able to align its goals with the outsourcing provider and encounter operational disruptions. Furthermore, the absence of key performance indicators (KPIs) and service level agreements (SLAs) can complicate the measurement of success, making it challenging to assess whether the outsourcing effort is meeting its intended outcomes (Lacity et al., 2010). In essence, meticulous planning and establishing performance metrics are pivotal to achieving the desired benefits and ensuring a seamless outsourcing transition.

- **Inadequate Communication:** Effective communication is the cornerstone of a successful outsourcing partnership. Inadequate communication between the organization and the outsourcing provider can result in various issues, from misunderstandings and missed deadlines to suboptimal outcomes (Kakabadse, & Kakabadse, 2000). Particularly in the case of offshore outsourcing, cultural and language barriers can compound these communication challenges, making it essential to bridge these divides effectively. Overcoming these barriers demands language proficiency, cultural sensitivity, and adaptability. By fostering transparent and open communication channels, organizations can mitigate these challenges and ensure that their outsourcing efforts are on track to meet their objectives (Quinn, & Hilmer, 1994).

- **Loss of Control:** The perceived loss of control is a common concern when organizations engage in outsourcing arrangements. It stems from delegating specific processes or functions to external partners, raising apprehensions about maintaining quality, security, and compliance standards (Lee, & Kim, 1999). This unease is often exacerbated by overly rigid contracts that leave little room for adaptability and responsiveness to changing circumstances. Striking a balance between delegating responsibilities and retaining oversight is crucial in addressing these concerns. Organizations can mitigate these issues by implementing transparent communication, well-defined service-level agreements, and a collaborative approach with outsourcing partners to maintain control while achieving the desired operational efficiencies (Kern, & Willcocks, 2000).

- **Hidden Costs:** While organizations often turn to outsourcing to achieve cost savings, it is crucial to be aware of hidden expenses that can arise throughout the outsourcing journey (Doh, & Teegen, 2002). These concealed costs may include transition expenses, legal fees associated with contract negotiations, and ongoing management expenses. Moreover, unexpected cost escalations from the outsourcing vendor can erode the anticipated savings, ultimately affecting the overall financial benefit of the outsourcing arrangement (Lacity, & Willcocks, 2001). To address this challenge, organizations should conduct comprehensive cost-benefit analyses, factor in all potential expenses, and establish precise cost controls and monitoring mechanisms to ensure that the outsourcing initiative remains financially viable and aligns with its strategic goals.

- **Quality Issues:** Quality issues can be a significant hurdle in outsourcing arrangements, potentially leading to subpar products or services that must meet organizational standards (Hindle, T., & Gudergan, 2015). This problem often stems from poorly defined quality standards, insufficient vendor performance, and a lack of effective oversight and quality control mechanisms. Organizations must establish clear and mutually agreed-upon quality expectations to mitigate these concerns within their service level agreements (SLAs). Additionally, implementing robust monitoring and performance evaluation processes and regular communication with the outsourcing partner can help ensure that quality remains a top priority throughout the outsourcing relationship, ultimately aligning the partnership with the organization's goals and standards (Bhagwat, & Sharma, 2007).

- **Security Risks:** Security risks are a significant concern in outsourcing, especially when sensitive data or valuable intellectual property is in play (Ang, & Inkpen, 2008). If the outsourcing provider does not have robust data protection measures and stringent cybersecurity practices, it can leave the organization vulnerable to potential breaches and data leaks. Ensuring that outsourcing partner adheres to the highest data security standards is paramount to safeguarding the organization's confidential information and intellectual assets. Security risks entail thorough vetting of the vendor's security protocols, clear contractual agreements regarding data protection (Lacity, & Hirschheim, 1993), and ongoing monitoring to maintain the integrity of your data throughout the outsourcing relationship.

- **Resistance from Internal Teams:** Resistance from internal teams is a common challenge regarding outsourcing initiatives. Employees concerned about potential job loss or decreased job security (Willcocks, & Kern, 1998) due to outsourcing can be understandably apprehensive. This resistance can manifest as decreased morale and productivity, creating hurdles to the smooth implementation of outsourcing efforts. These strategies should include effective change management strategies, transparent communication, opportunities for retraining or redeployment of affected employees, and demonstrating the long-term benefits of outsourcing to the organization and its workforce. By actively involving and reassuring employees, organizations can foster a more cooperative atmosphere and help ease the transition to outsourcing (Beulen, Ribbers, & Roos, 2008).

- **Inflexibility:** The need for adaptability in long-term outsourcing contracts can pose significant challenges for organizations. Such rigidity can hinder the organization's ability to respond effectively to shifting market dynamics or the rapid evolution of technology (Goo et al., 2009). In today's dynamic business environment, flexibility is paramount. Organizations need the agility to pivot, scale up or down, and embrace emerging technologies as they see fit (Carmel, & Nicholson, 2005). It is crucial for outsourcing contracts to incorporate provisions for adjustments and revisions, ensuring that the partnership remains aligned with evolving business needs and opportunities for innovation (Bhagwat, & Sharma, 2007).
- **Legal and Regulatory Compliance:** Maintaining legal and regulatory compliance is critical in outsourcing partnerships. Refraining from confirming that the outsourcing provider adheres to pertinent laws and regulations can expose the organization to many legal and reputational problems (Lacity, & Willcocks, 1998). Compliance shortfalls can trigger fines, lawsuits, and significant harm to the organization's reputation and brand. Organizations need to conduct rigorous due diligence and ongoing monitoring to ensure that their outsourcing partners operate within the boundaries of the law, mitigating these risks and safeguarding their integrity (Bapna, & Varadarajan, 2017).
- **Lack of Continuity Planning:** Organizations need more continuity planning in outsourcing. They might overlook essential contingencies, such as the possibility of the outsourcing vendor encountering difficulties or failing altogether. Without a solid backup plan, business continuity can be at risk (Rottman et al., 2013). To mitigate this, organizations should meticulously consider contingency scenarios and establish robust plans to ensure operations can continue smoothly despite unexpected vendor challenges, safeguarding their business interests and reputation (Kremic, Tukel, & Rom, 2006).

In analyzing the factors contributing to the failure in achieving organizational goals through IT outsourcing (Figure 4), our research identifies insufficient planning as the leading cause, with a criticality percentage of 94. This underscores the importance of comprehensive, strategic planning prior to engaging in outsourcing partnerships. Poor vendor selection is another paramount concern at 92%, which suggests that the success of outsourcing efforts is heavily dependent on the choice of the right service provider. Close behind, at 91%, is inadequate communication, indicating that clear, effective exchange of information between the company and its outsourcing partner is essential for meeting organizational objectives. Other significant challenges include hidden costs and security risks, rated at 88% and 84% respectively, pointing to the need for transparency in cost structures and stringent security protocols to protect sensitive data. Resistance from internal teams and inflexibility, with lower percentages, still represent considerable obstacles, hinting at the potential for internal pushback and a lack of adaptive processes. The lowest-rated concern, legal and regulatory compliance at 72%, reflects the necessity for outsourcing arrangements to be compliant with all applicable laws and regulations, a factor that, while less cited, remains crucial to the overall success and sustainability of outsourcing initiatives. These findings collectively illuminate the complexities and critical considerations that organizations must navigate to align IT outsourcing endeavors with their strategic goals.

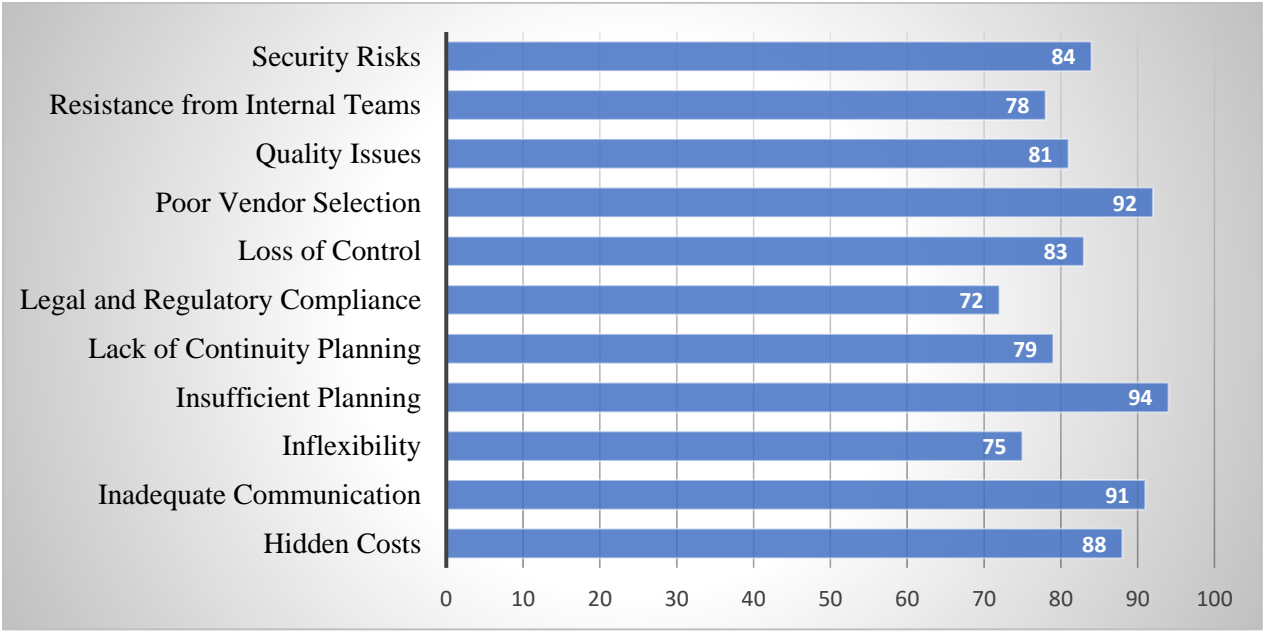

Figure 4. Reasons an Outsourcing Effort May Fail in Achieving Organizational Goals (Source: Survey on 100 respondents)

Organizations should invest in thorough planning, effective vendor selection, robust communication, and ongoing performance monitoring and management to mitigate these risks and increase the chances of a successful outsourcing effort. Additionally, they should remain flexible and adaptive in their outsourcing strategies to respond to changing circumstances.

**7. Discussion**

In our study, we discuss the critical elements that contribute to both the success and shortcomings of outsourcing initiatives. Our findings highlight that insufficient planning and poor vendor selection are the most prominent factors leading to the failure of achieving organizational goals, with percentages of 94% and 92%, respectively. This emphasizes the necessity for meticulous strategic planning and the careful vetting of potential outsourcing partners. Inadequate communication, rated at 91%, further underscores the need for a clear and effective exchange of information between all parties involved.

This study also explores additional challenges such as hidden costs and security risks, which are significant but often underestimated during the initial stages of outsourcing. These findings underscore the complexities that organizations must navigate to align their outsourcing endeavors with strategic goals. We conclude that organizations must systematically address these challenges, implement change management strategies, and ensure that contracts offer flexibility and robust compliance measures.

Outsourcing can be a valuable strategic tool when managed with awareness of potential pitfalls. By proactively addressing these challenges and employing best practices, organizations can harness the benefits of outsourcing, achieve their goals, and remain competitive in a rapidly evolving business environment. As we move forward, outsourcing is likely to remain a vital strategy for organizations seeking to thrive and innovate in an interconnected global landscape.

## 8. Conclusion

Outsourcing has become a prevalent strategy for organizations aiming to optimize operations, reduce costs, and focus on their core competencies in today's dynamic business landscape. While the benefits of outsourcing, such as cost savings, access to global talent, and scalability, are evident, organizations must also navigate potential pitfalls to ensure successful outcomes. The reasons for an outsourcing effort failing to achieve organizational goals underscore the critical importance of careful planning, strategic vendor selection, and effective communication. Vendor selection can lead to misaligning objectives, while adequate planning can result in operational disruptions and clarity of performance measurement. Communication can be improved, and a perceived loss of control can hinder adaptability. Hidden costs can erode anticipated savings, quality issues undermine organizational standards, and security risks pose significant threats. Resistance from internal teams, contract inflexibility, and compliance failures can further complicate outsourcing endeavors. Organizations must address these challenges systematically, implement change management strategies, and ensure that contracts incorporate flexibility and robust compliance measures. Outsourcing can be a valuable tool when managed strategically and with a keen awareness of potential pitfalls. By proactively addressing these challenges and leveraging best practices, organizations can harness the benefits of outsourcing while minimizing risks, achieving their organizational goals, and remaining competitive in a rapidly evolving business environment. As the business landscape grows, outsourcing will likely remain a vital strategy for organizations seeking to thrive and innovate in an increasingly interconnected world.